# Thermal motions in complex liquids: the *2D* Lennard-Jones liquid

Alexander Z. Patashinski[*a)], Rafal Orlik[b)], and Mark A. Ratner [a)]

a)  Northwestern University, Department of Chemistry, Evanston, IL 60208

b)  Orlik Software, Wroclaw, Poland

*Corresponding author: a-patashinski@northwestern.edu

**Abstract**

Collective thermal motions are studied in an aggregated *2D* Lennard-Jones liquid at thermal equilibrium and under shear flow. As a means to resolve different temporal and spatial scales, smoothing over increasing times is used. On times of few to few tens of particles vibration periods, particles in-cage vibrations and highest frequency longitudinal and transverse Hypersound dominate the picture. On times up to many thousands of particle vibration periods, the liquid appears spatially heterogeneous. On these times, non-oscillatory currents involving many particles manifest the hierarchical dynamics of the heterogeneity. These currents result in slow changes in temperature, density, and velocity profiles across the system persisting for surprisingly long times. Heterogeneity fades, and a crossover to non-fluctuational Hydrodynamics is observed only for smoothing times approaching many tens of thousands vibration periods. On these asymptotically-large time-scales, the liquid is spatially homogeneous in the bulk; in thin layers near the boundaries the degree of crystallinity increases and the mobility decreases due to liquid-boundary interactions.

## 1. Introduction

In a body, molecules randomly move with average thermal velocity [1, 2] $v_T=(Dk_BT/m)^{1/2}$ where $D=1, 2, 3$ is the number of spatial dimensions, $T$ the temperature, $m$ the molecular mass, and $k_B$ the Boltzmann constant. The direction of th5e actual velocity of a molecule changes due to interaction with other molecules; the characteristic time $\tau$ of this change serves as a natural time unit for thermal fluctuations in the system. Molecular interactions lead to correlations between molecules positions, and also their velocities. A known example of correlated, cooperative motion is sound. In sound motions, the center-of-mass velocity, density, and pressure and shear stress in a small volume oscillate with periods proportional to the wavelength $\lambda$ of the soundwave. On larger times, especially in complex liquids ("soft-matter") [1, 3-11], one also observes non-oscillatory collective motions. Visco-elastic effects in liquids can be explained by assuming that molecules form transient [7-10] aggregates. Then, on time-scales defined by the lifetimes of these aggregates, the liquid is heterogeneous [9], with the spatial scale of this heterogeneity determined by the size of the aggregates. Direct observation and measurement of aggregation dynamics is a challenging experimental task, so the current picture of the fluctuating local structure in complex liquids lacks important details.

An alternative possibility of studying microscopic-scale motions in a liquid is provided by Molecular Dynamics (*MD*) simulations in combination with specially devised visualization, recognition, and measurement programs. Here, we use this approach to study fluctuations in a two-dimensional (*2D*) Lennard-Jones (*LJ*) liquid. The choice of the model takes into account that, in a certain sense, the *2D* liquid near solidification represents the simplest, and easiest to study, complex liquid: previous studies [5, 7-10] found that at temperatures close to solidification, particles of this liquid aggregate into transient crystalline-ordered clusters ("crystallites") intertwined with less-ordered amorphous clusters. This dynamic mosaic of crystallites and amorphous regions appears statistically-stable, and bears some general resemblance to much more complicated complex *3D* liquids [6, 11].

In recent decades, *2D* systems have been intensely studied to test the nature of *2D* melting [13-16]. The results of intensive computer simulations generally confirm the defect-unbinding nature of melting, although close to the triple point there are some subtle violations of this scenario [14]. Here, we study the liquid at temperatures above melting, in states where the correlation length for the local order is a fraction of the system size so the liquid is locally ordered but globally disordered. At these temperatures, a particle in the liquid is surrounded by several (4-6) close neighbors (referred to as cage particles), and vibrates in this environment with the average period $\tau$. Unlike the case of a crystal where a particle vibrates in same cage for macroscopically large times, in a liquid the lifetime $\tau_r$ of the near environment of a particle (the cage time [17]) is microscopic; this time varies substantially depending on thermodynamic parameters. In simple *2D* and *3D* liquids [2, 9], this lifetime is of the order of $\tau$. In the *2D* Lennard-Jones liquid, $\tau_r \sim \tau$ at a temperature $T$ that is ~50% above the melting temperature, increases one-two orders of magnitude on cooling to the aggregated state studied here, and becomes large beyond direct observation on further cooling/compression to crystalline state.

To perturb the thermodynamic equilibrium, the liquid in our study is placed between solid boundaries moving in opposite directions (see Fig. 1). For a macroscopic system, similar conditions generate the Couette flow [18, 19]. The current particles configuration is visualized, and instant (at a current iteration step) or smoothed (time-averaged) profiles of the local temperature $T(r,)$ density $\rho(r)$, and velocity $v(r)$ across the system are studied. For sufficiently large smoothing times $\theta > \theta_H$, these profiles become (see Fig. 2) the non-fluctuating, smooth profiles predicted by the Hydrodynamics [18, 19], but for smaller smoothing times $\theta < \theta_H$, the profiles fluctuate in space and time (Fig. 3-6).

Factors that make the aggregated state of the *2D* liquid stable are specific for *2D* liquids: in *3D* liquids, significant aggregation is observed or assumed only in systems with complicated and competing interactions. In quasi-*2D* liquid of hard and soft spheres confined between two parallel planes, aggregation diminishes with increasing distance between these planes [7, 8]. We believe that general laws of complex behavior are common for complex liquids, so studies of the relatively simple *2D* liquids can give an insight into the internal dynamics of structurally much more complicated *3D* liquids.

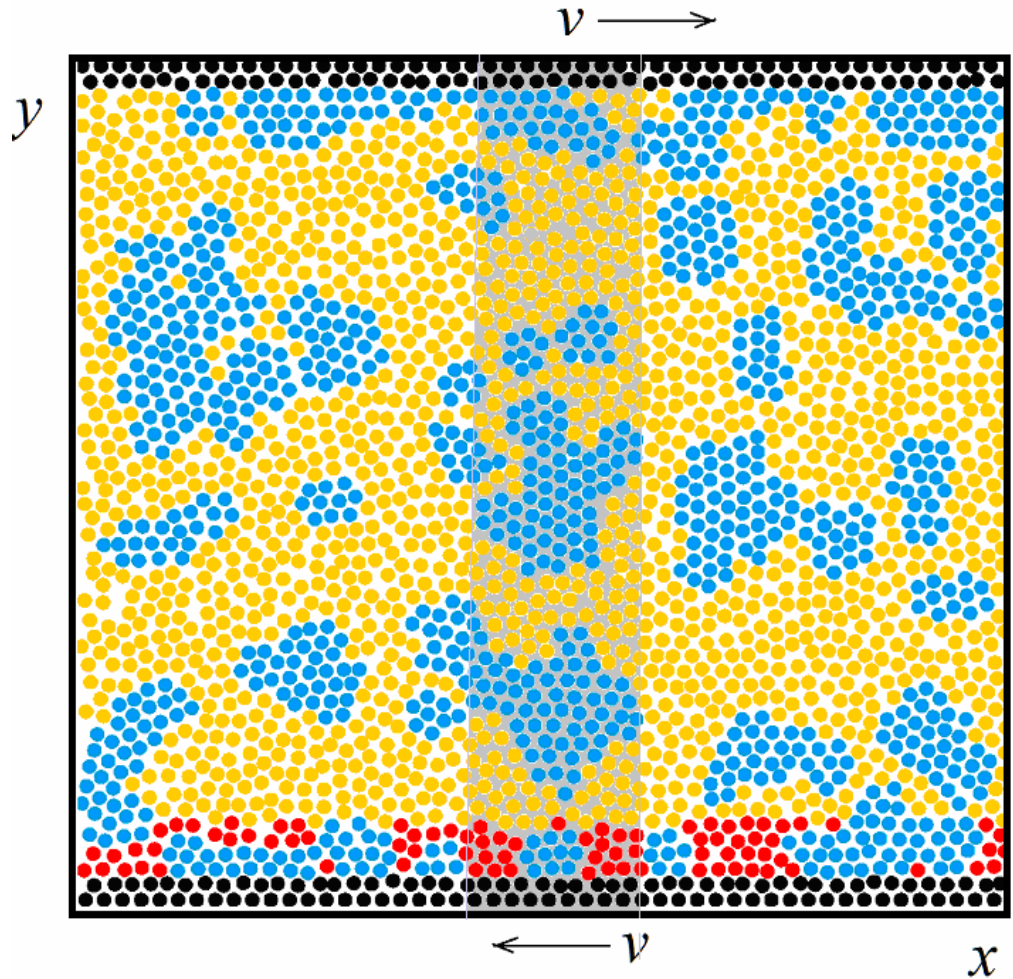

ig. 1. A system of *N*=2500 particles at *T*=2, $\rho$=0.92, *v*=0.01. Particles in the temperature-monitored part are colored red, particles in some crystallites-blue, boundary particles - black; the measuring column is indicated by a gray background.

## 2. The system

Molecular Dynamics (*MD*) with Verlet velocity algorithm is applied to simulate *2D* systems of *N*=2500÷62500 Lennard-Jones particles in periodic boundary conditions; this rather standard algorithm is described in many publications – see, for example, [9, 10, 13-15, 20, 21]. We use a Euclidean coordinate system with the *x*-axis parallel and the *y*- axis orthogonal to the boundaries; due to periodicity along the x-direction, the system has the geometry of a cylindrical shell limited by the top and bottom boundaries (see Fig. 1). To create solid boundaries, special Lennard-Jones particles (black circles in Fig.1) are placed at the bottom ($y=0$) and the top ($y=L_y$) of the simulation cell; the linear density of these boundary particles is chosen sufficiently large to prevent particles of the liquid from penetrating the boundaries. Boundaries move as solid bodies, parallel to the x-axis but in opposite directions, with velocities $v$ for the top and $-v$ for the bottom boundary.

The Lennard-Jones interaction potential is taken in the standard form

$$U(r) = 4\varepsilon \left[ \left(\frac{\sigma}{r}\right)^{12} - \left(\frac{\sigma}{r}\right)^{6} \right], \qquad 1$$

where $r$ is the distance between interacting particles, $\sigma$ the Lennard-Jones size of a particle, and $\varepsilon$ the characteristic Lennard-Jones energy. In simulations, the potential (1) is truncated at the cut-off distance $r_{tr}=3\sigma$. Below, the results of simulations are described using reduced Lennard-Jones units [20, 21, 9, 10]; in these units, $\sigma=1$ and $\varepsilon=1$, the reduced temperature is $T=T^*/(\varepsilon k_B)$ where $T^*$ is the temperature in "physical" units and $k_B$ the Boltzmann constant. The reduced density is $\rho=\sigma^2\rho^*$ where $\rho^*$ is the particle number density in "physical" units. The Lennard-Jones time $\tau=\varepsilon/(m\sigma^2)^{1/2}$, of the order of particles vibration period, is chosen as the reduced time unit; the integration step is defined as $h=1/500\,\tau$. We found no significant changes in the behavior when using smaller integration steps ($1/1000$-$1/32000\,\tau$).

The temperature in the liquid is stabilized by applying the standard *NVT* algorithm [20] to a small (typically 1/10) part of the liquid (see red region near the bottom boundary in Fig. 1). In the rest of the system, the temperature is determined by the balance between heat release due to velocity gradients and heat exchange with the *T*-maintained layer. This way of temperature maintenance is chosen to mostly confine to a small layer the perturbations of particles "natural" motion by the (*NVT*) temperature maintenance procedure. For steady flows with average velocity gradients $<dv_y/dy>=2v/L<10^{-4}$, the smoothed temperature profile is linear within 2% accuracy. For larger gradients, we observed remarkable nonlinear regimes including phase separation and appearance of crystalline or gaseous sub-systems; although interesting, these regimes are beyond the scope of the current paper.

To facilitate data collection, the system is partitioned into measuring cells by *n* equidistant horizontal and *n'* equidistant vertical lines $x=nL_x/s$, $y=n'L_y/s'$ (*n* =1, ... , *s*-1; *n'*=1, ... , *s'*-1). Here, $L_x$ and $L_y$ are the length and the width of the system (see Fig. 1). The dimensions of a measuring cell are then $\Delta x= L_y/s'$ and $\Delta y= L_x/s$; the number $\Delta N(r,t)$ of particles in a cell determines the density $\rho(r,t)=\Delta N(r,t)/\Delta x\Delta y$. Instant (at the currant iteration step) profiles $X(y,t)=[T(y,t), \rho(y,t), v_x(y,t), v_y(y,t)]$ are defined as cell-averaged values of corresponding characteristics in vertical column of measuring cells (gray background in Fig.1); the coordinate *y* refers to the cell center. The profiles $X(x,t)=[T(x,t), \rho(x,t), v_x(x,t), v_y(x,t)]$ give the data for a horizontal row of measuring cells. Smoothed profiles $<X(y,t)>_\theta$ , $<X(x,t)>_\theta$ are obtained by smoothing (time-averaging) instant characteristics over a time interval (*t-θ/2, t+θ/2*).

When the density increases along the *T*=2 isotherm, the fraction of particles in crystalline-ordered clusters grows from zero at $\rho<0.65$ to ~10% at $\rho=0.87$, ~20% at $\rho$=0.90, and ~40% at $\rho$=0.92 (see Fig. 2 where particles in some large crystallites are painted gray). At $T$=2, $\rho$=0.92, crystallites are intertwined by amorphous clusters. In amorphous clusters, a particle makes frequent diffusion jumps, changing near neighbors on times ~$\tau$ [9]; this feature is typical for non-aggregated states at $\rho$<0.65. On increase of the density or decrease of the temperature, crystallites percolate and merge into a multi-connected crystalline matrix [9, 10]. The liquid at $T$=2, $\rho$=0.92 is already close to but still not in this percolated state; unless stated otherwise, the data and discussion below refer to this non-percolated state.

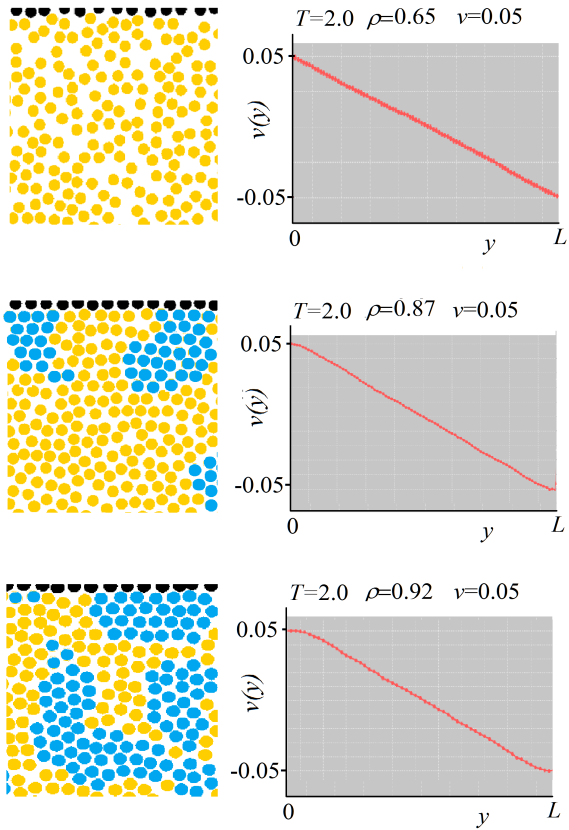

Fig. 2. The structure and the asymptotic velocity at $T=2$ for $\rho=0.65$, 0.87, and 0.92.

**3. Steady states, and crossover to Hydrodynamics**

A sufficiently long simulation under steady $(T\rho v)$-parameters (temperature $T$, density $\rho$, boundary velocity $v$) brings the system close to a (statistical) steady state; the time necessary to essentially "forget" the initial conditions and reach a "practical" steady state depends on system size and initial conditions, it is $\sim 10^4 \tau$ for a system of $N=40000$ particles at $T=2$, $\rho=0.92$ and crystalline initial configuration. As a practical definition, we identify a state as steady when, with the accuracy of 2% of short-time fluctuations, ***a***) the profiles $<X(y,t)>_\theta$ and $<X(x,t)>_\theta$ remain unchanged, and ***b***) these profiles are universally reproducible in repeating simulations. The smoothing time $\theta_r$ necessary for the steady-state fluctuations to fit into the 2% limit depends on temperature and density (see below); for $T=2$, $\rho=0.92$, and $v/L=10^{-4}$, $\theta_r \sim 10^4$. Reproducibility and uniformity of the smoothed profiles indicate that the time $\theta_r$ is sufficient for the system to sample a representative ensemble of configurations and local structures. On these asymptotic time-scales, the liquid can be treated as a continuum medium, and Hydrodynamics [18] can be used to describe large scale quasistatic motions in the liquid.

Fig. 2 presents asymptotic steady-state profiles $v_x(y)=<v_x(y,t)>_\theta$ for liquids at $T=2$, $v=0.05$, and $\rho=0.65$, 0.87, 0.92. The smoothing time in these plots is $\theta=10^5 \tau$; fluctuations in these profiles are well within the 2% range. Approach to the asymptotic time-scale, and the decrease of fluctuation amplitudes with increasing smoothing time is discussed in the next Section.

In accordance with the predictions of Hydrodynamics, the asymptotic velocity $v_x(y)$ in the middle part of the system is a linear function of the coordinate $y$: $\delta v_x(y)/\delta y = 2v/L = const$. For $\rho=0.87$ and $\rho=0.92$ one finds in Fig. 2 layers of decreased mobility (and increased crystallinity) near the boundaries. Those layers become detectable for $\rho>0.70$ but remain relatively thin for $0.70<\rho<0.92$. Upon further increase of the density, crystalline boundary layers rapidly widen, and act as extensions of the solid boundaries, until the top and the bottom layers overlap and the entire liquid becomes crystalline with the flow facilitated by dislocations. For small average velocity gradients $v/L<10^{-4}$, the width of the layer does not depend on the velocity. One assumes that the boundary layers signal increased crystallinity due to the interaction of the liquid with the crystalline boundary; the slop of the asymptotic profile $v_x(y)$ serves then as a measure (somewhat non-local) of the average degree of crystallinity at a given distance from boundaries.

## 4. Collective thermal motions in the liquid

In the liquid at $T=2$, $\rho=0.92$, the number of particles in a large crystallite is $\delta N \sim 20\text{-}70$. Connecting each internal (not at the border) crystallite particle with its cage particles creates a near-neighbors network that can be mapped onto a finite-size hexagonal crystal [9]. Borders of a crystallite visibly fluctuate in time due to small (3-5 particles) random increases or decreases of the crystallite at the border [10]. As a result of these micro-melting/micro-crystallization events, the crystallite (as a finite-size network) changes its shape and position without substantial changes in the positions of internal particles. At $T=2$, $\rho=0.92$ the average time for a particle to continuously remain in a crystallite is ~$15\tau$. Direct observation proves that density differences between crystalline-ordered clusters are substantially smaller than the difference (~15%) between a crystalline and amorphous cluster.

Spatial homogeneity, illustrated by Fig. 2, appears only in characteristics smoothed over asymptotically-long times $\theta > \theta_r$ on which the system samples a representative ensemble of local structures. Smoothing over sub-asymptotic time intervals $\theta < \theta_r$ does not completely remove heterogeneity: the sub-ensembles sampled on these times are constrained by slowly changing characteristics of the local structure. Changes of this structure are hierarchical. Below, we track these changes on progressively increasing time-scales.

*a. Density fluctuations*

The instant (at the current iteration step) number $\Delta N(t)$ of particles in a fixed-size measuring cell is fluctuating, as illustrated by Fig. 3 presenting a short-time ($\Delta t=500h=\tau$) sampling of the density $\rho(t)=\Delta N(t)/(\Delta x \Delta y)$ in a small square-shaped cell with spatial dimensions $\Delta x=\Delta y=5$. The density changes in elementary steps (easily recognizable in Fig. 3) that are particle crossings of the cell border. Each single crossing changes the density by ~5%. Most events in Fig. 3 are generated by particles vibrating near the borders of the cell; for the 5X5 cell, the average frequency of these events is $\nu_0 \sim 10$. The resulting high-frequency noise can be diminished by smoothing the data over few particle vibration periods, or increasing the cell size (see Fig. 5). A slow contraction (or expansion) of the liquid in the cell changes the amount of time spent by a near-border particle inside (and outside) the cell, and manifest itself in density changes averaged over times $\theta > 1/v_0$. Periodic contractions/expansions with frequencies $v < v_0$ (Hypersound waves) are described in the next Section.

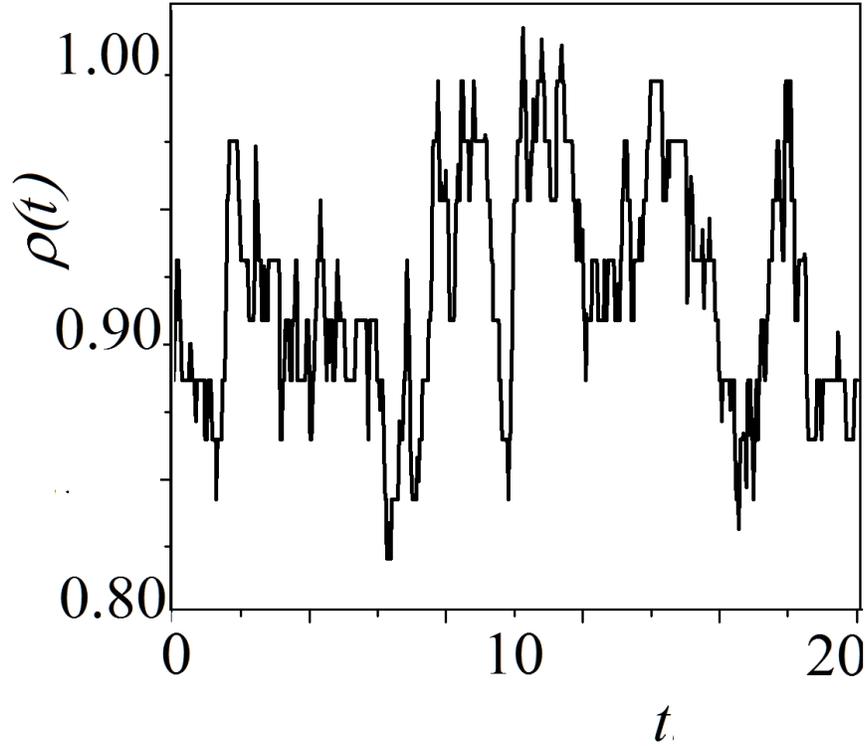

Fig. 3. Density oscillations in a 5x5 cell; sampling time $\Delta t=20$.

Direct observation reveals that the volume fraction of crystalline clusters in the small measuring cell fluctuates between close to 100% (when the entire cell is occupied by a crystallite) and 0% (when the cell is occupied by an amorphous cluster). The density then fluctuates between that of a crystal and of an amorphous cluster; the expected (and observed) amplitude of these fluctuations is of the order of density difference (~15%) between a crystallite and an amorphous cluster. In a small cell, this particles rearrangement mechanism gives the largest contribution to density fluctuations. A possible contribution of fluctuating currents (see next sub-Section) bringing differently-ordered clusters into the cell appears small due to the very small thermal displacements of particles caused by these currents.

Fig. 4 presents $\rho(t)$ sampled in the same small cell but over $\Delta t=2000\tau$ ($10^6 h$). The density is smoothed (by adjacent averaging) over $\theta=0.2$ (100$h$, the black line), $\theta=10$ (5000$h$, the red line), and over $\theta=100$ (50000$h$, the cyan line). The red-line peaks have an average period ~10-20, about the lifetime of a crystallite, and describe random crystalline-amorphous switches in the measuring cell, with density changing ±5%. The cyan line variations have smaller amplitude (±0.5%), and resemble oscillations with a period ~1/200; we suggest that these variations represent a low-frequency Hypersound mode.

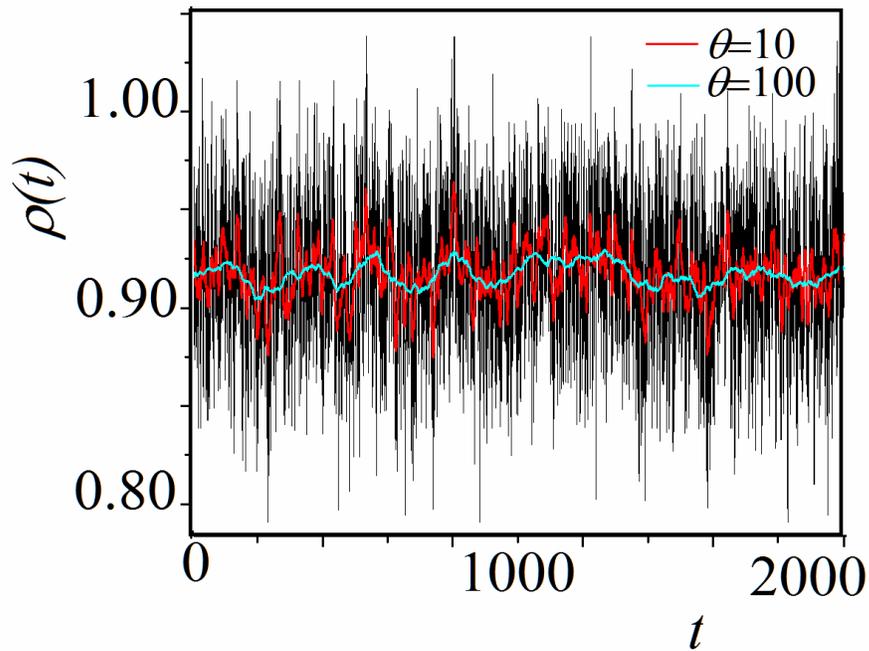

Fig. 4. $\Delta t$=2000 ($10^6 h$) sampling of the density. Smoothing times: 0.2 (100$h$, black line), $\theta$=10 (5000$h$, red line), and $\theta$=100 (50000$h$, cyan line).

The *FFT* (*Fast Fourier Transform*) power spectrum of the difference $\delta\rho=\rho(t)-<\rho>$ is shown in Fig. 5. The frequency range $\nu$>0.05 of this spectrum is discussed in the next Section. In the range $\nu$<0.01, density fluctuations are, roughly, frequency-independent, indicating that structure changes at corresponding long times are uncorrelated. Spectra for even longer sampling times ($\Delta t$=10000) and/or larger (10X10) measuring cells support this assumption.

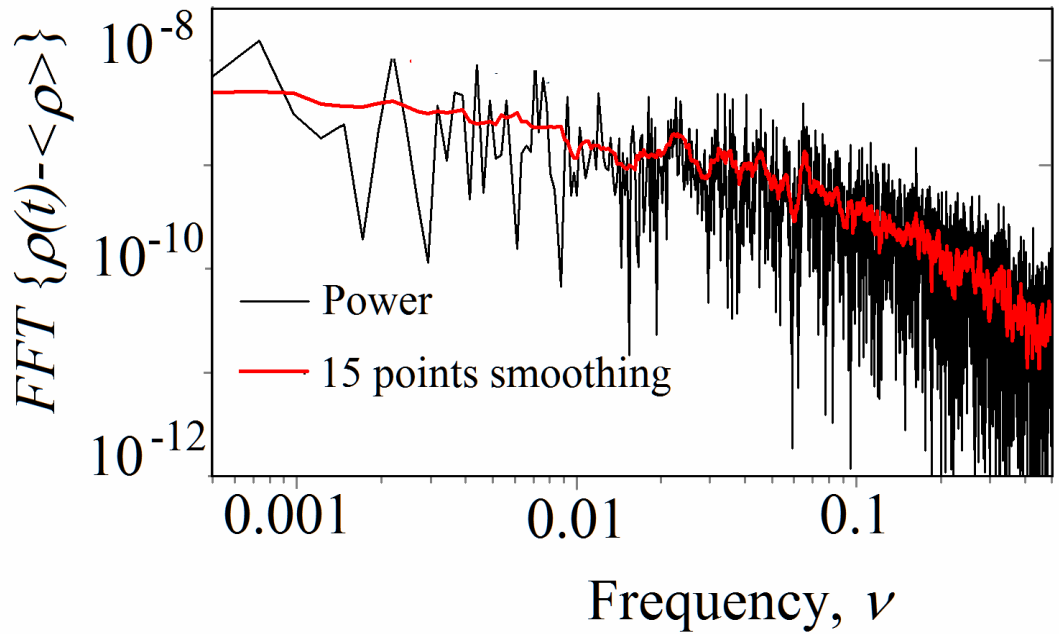

Fig. 5. *FFT* (*Fast Fourier Transform*) of density fluctuations shown in Fig. 4.

## b. Thermal mixing and fluctuating currents

In a complex liquid, the trajectory of a single particle reveals a complex hierarchy of thermal motions [9], with the particle intermittently trapped in crystallites for significant time. Motions of single crystallites are correlated into fluctuating currents. We study these currants by smoothing the velocity profiles over progressively increasing times: smoothing a profile over a time $\theta$ decreases (as $(\theta \nu)^{-1}$) the amplitudes of fluctuations with frequencies $\nu > 1/\theta$ while not changing amplitudes for $\nu < 1/\theta$.

The asymptotic velocity $v_x(y)$ in Fig. 2 describes the steady, non-fluctuating shear flow in the liquid, an average over very large times. The difference $\delta<v_x(y,t)>_\theta = <v_x(y,t)>_\theta - v_x(y)$ describes the fluctuating currents in the liquid at a time-scale $\theta$ (Fig. 6). For $\theta = 100$ (the black line in Fig.6), one finds regions (blocks) of a size $\Delta R \sim 40$ moving with velocities $\delta v_x \sim 0.01$ relative to the background. The relative displacement of a block during this time is then $\theta \delta v_x \sim 1$. The random relative velocities of smaller ($\Delta R \sim 10$, the size of large crystallites) regions are of same order. There are no smaller-size details in the plot, although the measuring cells in Fig. 6 have $\Delta y = 4$.

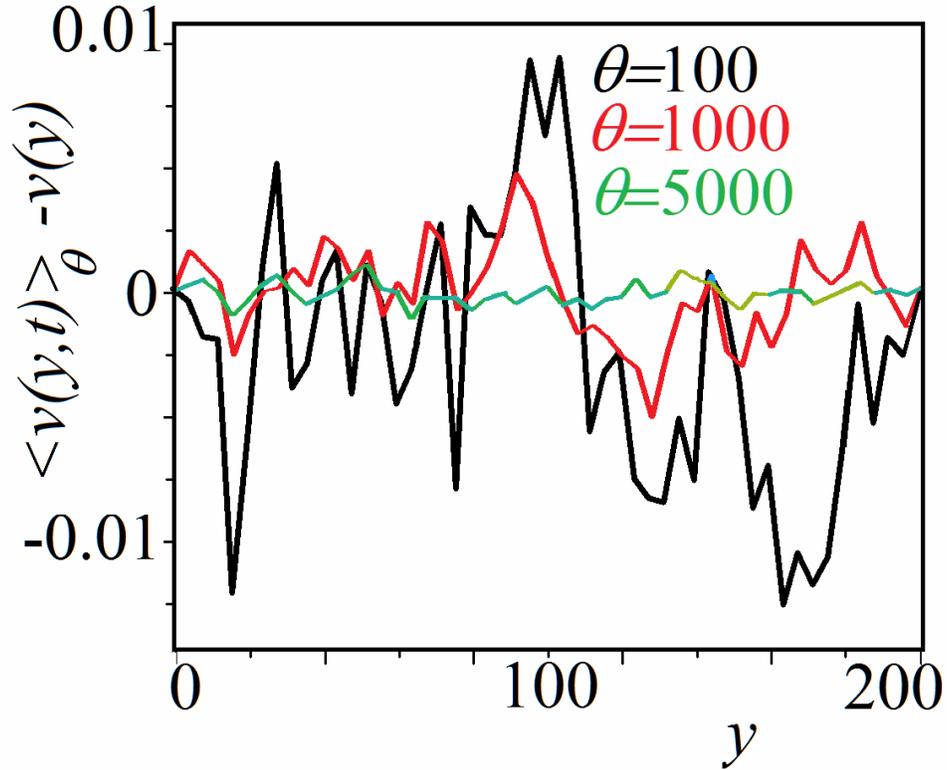

Fig. 6. Changes in $\delta<v_x(y,t)>_\theta$ with increasing smoothing time $\theta$.

On a larger time-scale $\theta=1000$ (the red line in Fig. 6), the average velocities both of the $\Delta R \sim 40$ blocks and the $\Delta R \sim 10$ regions are 2-4 times smaller than for $\theta=100$, quantitatively consistent with the prediction of the random walk theory. However, for $\theta=5000$ (the green line), all displacements decreased faster than this theory predicts. We suggest that for times of hundreds and few thousands of vibration periods, the viscoelastic liquid retains substantial elasticity that prevents large thermal displacements. An independent proof of this short-times elasticity give the transverse Hypersound oscillations with periods of at least few hundreds of particles vibration periods – see the next Section.

Thermal currents are driven by thermal fluctuations. In a non-equilibrium steady state with an average velocity gradient $dv(y)/dy=2v/L$, fluctuations of the local mobility may although lead to velocity fluctuations (proportional to the average gradient). In particular, fluctuations of mobility may reflect fluctuations in the ratio of the crystalline to amorphous component –relations between this ratio and liquid mobility explain the appearance of the boundary layers in Fig. 3. To estimate the possible contributions of mobility fluctuations, we plotted (see Fig. 7) the profiles $\delta\!<\!v_x(y,t)\!>_\theta$ for $N=40000$, $T=2$, $\rho=0.92$, $\theta=1000$ for an equilibrium system ($v=0$, the blue line) and for a non-equilibrium liquid at $v=0.01$ (the red line). Both profiles exhibit the same level of fluctuations. One concludes from Fig. 7 and similar plots that the mobility fluctuations contribution in a non-equilibrium steady state are at most of same order but most likely much smaller than the thermally-induced velocity fluctuations. The same can be stated about other fluctuations: within the accuracy of our observations, fluctuations in weakly-nonequilibrium steady states are not changed by the non-equilibrium. However, the accuracy of our observations is insufficient for a quantitative statement: a more quantitative approach assumes comparing representative *ensembles* of configurations, a task that is beyond the scopes of this study.

In addition to $<\!v_x(y,t)\!>_\theta$, we also analyzed the profiles $<\!v_y(y,t)\!>_\theta$, and also the profiles $<\!v_x(x,t)\!>_\theta$ and $<\!v_y(x,t)\!>_\theta$ obtained using a row of measuring cells stretched along the *x*-axis. All these profiles give the same picture of slow thermal currents and decaying correlations between particles velocities on times $\theta>1000$.

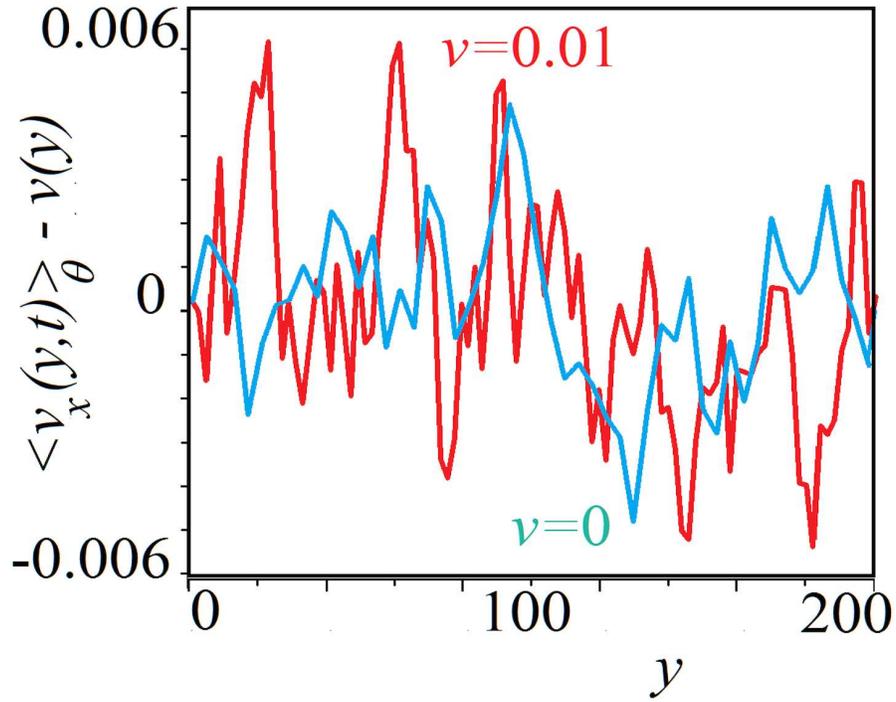

Fig. 7. Comparison of equilibrium ($v=0$, blue line) and non-equilibrium ($v=0.01$, red line) velocity $\delta \langle v_x(y,t) \rangle_\theta$ fluctuations.

*c. Hypersound*

At the high-frequency end of the sound spectrum, the periods of soundwaves approach the particles vibration period; the corresponding wavelengths are close to intermolecular distances. As discussed above, on these short times, the liquid is acoustically heterogeneous. Correspondingly presents the velocity profiles $\langle v_y(y,t) \rangle_\theta$ (smoothed over $\theta=1$) for two sampling intervals: $(t, t+1)$ and $(t+3, t+4)$. Each profile consists of alternating regions of positive and negative velocity gradients indicating compression or expansion of the liquid; the average distance between neighboring velocity maxima is $\Delta R \sim 10$. The shift time $\Delta t=3$ is enough for some velocity maxima to become velocity minima (and vice versa), a sign that there are oscillations with a period $2\Delta t=6$. The yellow/blue regions mark regions where the velocity increased/decreased during the shift time. Yellow regions are regularly intertwined by blue regions, with the distance between one-color regions $\lambda \approx 25=200/8$. The red/blue alternations form a wave-like pattern roughly resembling an eigenmode of an acoustic resonator of the form $X(y,t)= \sin(2\pi\nu t)\sin(2\kappa\pi y/200)$ with $\kappa=8$, $\nu \approx 0.15$.

An eigenmode $m$ of a linear resonator is an oscillator characterized by eigenfrequency $\nu_m$, the mass $M_m \sim \rho V$ associated with the mode, and the quality ($Q$) factor $Q_m$. When excited by the thermal bath, this oscillator has the thermal noise power spectrum

$$S_m(\nu) = \frac{2kT}{(2\pi)^4 M_m Q_m} \frac{\nu_m}{(\nu_m^2 - \nu^2)^2 + (\nu\nu_m/Q_m)^2}. \qquad 2$$

For $Q_m \gg 1$, this spectrum has a sharp peak at $\nu \approx \nu_m$; at the peak $S_{peak} \sim (Q_m/M_m) \nu_m^{-3}$. A time-independent acoustic heterogeneity may complicate the spatial picture of an eigenmode. In the case studied here, the heterogeneity is dynamic: the positions of crystallites borders substantially change on times $\theta \sim 20$, while crystallites motions and positions are correlated for times $\theta \sim 1000$. For a dynamically heterogeneous medium, description of a resonator in terms of eigenmodes is at best qualitative.

The small ($L \approx 200$) size of the resonator puts an upper limit on acoustic eigenfrequencies. For a rough estimate of this upper limit, one assumes that the period $1/\nu$ of a mode oscillations is proportional to the characteristic wavelength $\lambda(\nu) \sim 1/\kappa$. For the assumed mode $\kappa=8$ (see Fig. 8), the frequency $\nu\approx0.1$; then, the mode $\kappa=1$ is expected to have the frequency $\nu_{min}\approx0.03$, near the end of the power-low-like behavior of $S(\nu)$ in Fig.5. Below $\nu=0.01$, $S(\nu)$ only weakly depends on frequency; this part of the spectrum represents aperiodic motions related to hierarchical changes in the structure of the liquid.

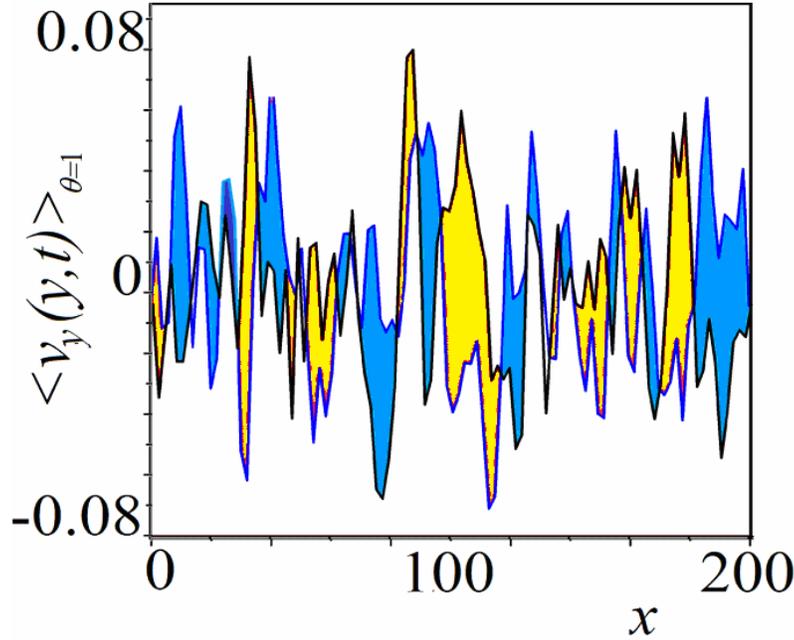

Fig. 8 . Velocity profiles $<v_y(y,t)>_{\theta=1}$ at times $t$ and $t+3$. The blue (yellow) regions mark velocity increase (decrease).

The decrease of fluctuation amplitudes with increasing smoothing time can be used to roughly estimate the wavelength $\lambda(\nu)$ to frequency $\nu$ relations for the sound modes. Fig.9 presents *FFT transforms* of the smoothed velocity $<v_y(y,t)>_\theta$ (as function of $y$) for $\theta=1,10$, and 20. As the plot shows, the correspondence is rather loose, but the peaks indicating modes are distinguishable, especially for lower frequencies (and wave-vectors).

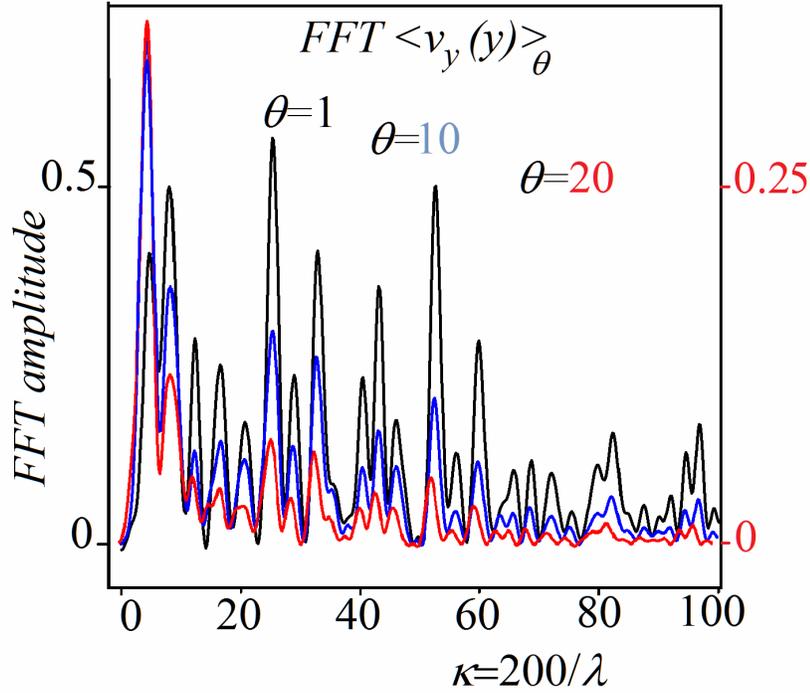

Fig. 9. *FFT*-amplitudes of smoothed velocity profiles for different smoothing times, $\theta$. Note the change of units for the red line.

The profile $v_y(y,t)$ describes particles longitudinal (along the measuring column) motions. Similarly, $v_x(x,t)$, describes longitudinal motions along a measuring row oriented parallel to the *x*-axis. Motions of particles in the direction orthogonal to the measuring column or row are described by the transverse profiles $v_x(y,t)$ and $v_y(x,t)$. At $T=2$, $\rho=0.92$, the liquid appears elastic for $\theta<100$. On times where elasticity prevails, the liquid behaves as a non-uniform crystal. The overall picture of standing-wave-like transverse oscillations with frequencies $v<1/100$ observed using the transverse profiles $v_x(y,t)$ and $v_y(x,t)$ is qualitatively similar to the longitudinal fluctuations shown in Fig.8 and 9.

## 5. Discussion and conclusions

We presented a rather detailed picture of internal motions in an aggregated *2D* liquid. On time-scales of up to few thousand of particles vibration, the liquid is heterogeneous and represent a mosaic of ordered and less-ordered clusters. There is increasing evidence [11, 12] that this type of heterogeneity is a common feature of condensed but not long-range ordered systems. In some well-studied liquids, a local order resembling that in a crystal has been demonstrated long-ago. A known example here is liquid Benzene where some elements of a rather complex crystalline local arrangement have been found in X-ray studies [25]. We suggest that liquid-liquid phase transitions [26-29] are rather dramatic manifestations of this kind of the local order, and assume that a mosaic of ordered and disordered clusters is a common feature of locally-ordered but globally-disordered systems, and that molecules redistribution between ordered and disordered (or less-ordered) regions determines the unusual properties of these systems.

The simplest and easiest to study complex, aggregated liquids are the *2D*-liquid near its crystallization temperature; these systems, in particular the *2D* Lennard-Jones liquid, can be suggested as a test-bed for studies of the dynamics of particles aggregation, aggregates clustering, and the interaction of local heterogeneities with externally-imposed flows.

*Acknowledgements:* We thank Monica Olvera de la Cruz and Creighton Thomas for help in the early stages of this work. **This work was supported by Center for Bio-**

**Inspired Energy Science (CBES) which is an Energy Frontier Research Center funded by the U.S. Department of Energy, Office of Science, Office of Basic Energy Sciences under Award No. DE-SC0000989.****References**

1. J.-L. Barrat and J.-P. Hansen, *Basic concepts for simple and complex liquids* (Cambridge University Press, New York 2003).

2. J-P. Hansen, I.R. McDonald , *Theory of Simple Liquids, 4th Edition* (Academic Press, 2013).

3. R.G. Larson, *The Structure and Rheology of Complex Fluid.  Topics in Chemical Engineering* (Oxford University Press, New-York-Oxford, 1999).

4. G.R. Fleming and P.G. Wolynes, Physics Today  **43**(5), 36 (1990).

5. X. Xu, St. A. Rice, and A. R. Dinner, PNAS 110, 3771-3776 (2013).

6.  M.D. Ediger, Ann. Rev. Phys. Chem. **51**, 99 (2000).

7. A. S.-Y. Sheu and St. Rice, J. Chem. Phys. 128, 244517-1-8 (2008).

8. A. S.-Y. Sheu and St. Rice, J. Chem. Phys. 129, 124511-1-11 (2008).

9. A.Z. Patashinski, R. Orlik, A.C. Mitus, B.A. Grzybowski, and M.A. Ratner, J. Phys. Chem. C  **114**, 20749 (2010).

10. A.Z. Patashinski, M.A. Ratner, B.A. Grzybowski, R. Orlik, A.C. Mitus, J. Phys. Chem. Lett. **3**, 2431 (2012).

11. C.A. Angell, MRS Bull. **33**, 544 (2008).

12. K.J. Strandburg, Rev. Mod. Phys. **60**, 161 (1988).

13. K. Chen, T. Kaplan and M. Mostroller, Phys. Rev. Lett. **74**, 4019 (1995).23

**Figures and captions:**

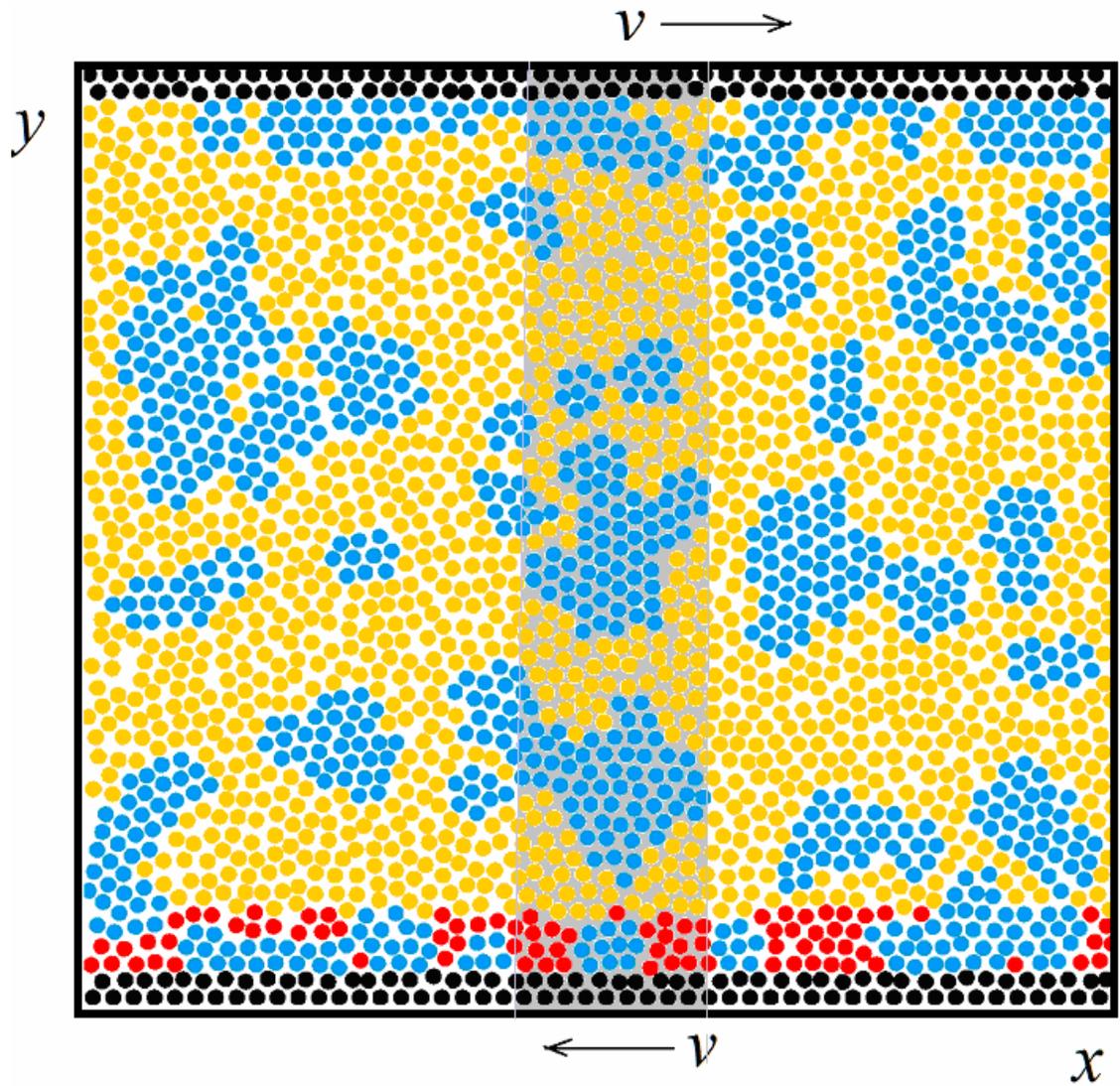

Fig. 1 (color online). A system of $N$=2500 particles at $T$=2, $\rho$=0.92, $v$=0.01. Particles in the temperature-monitored part are colored red, particles in some crystallites-blue, boundary particles - black; the measuring column is indicated by a gray background.

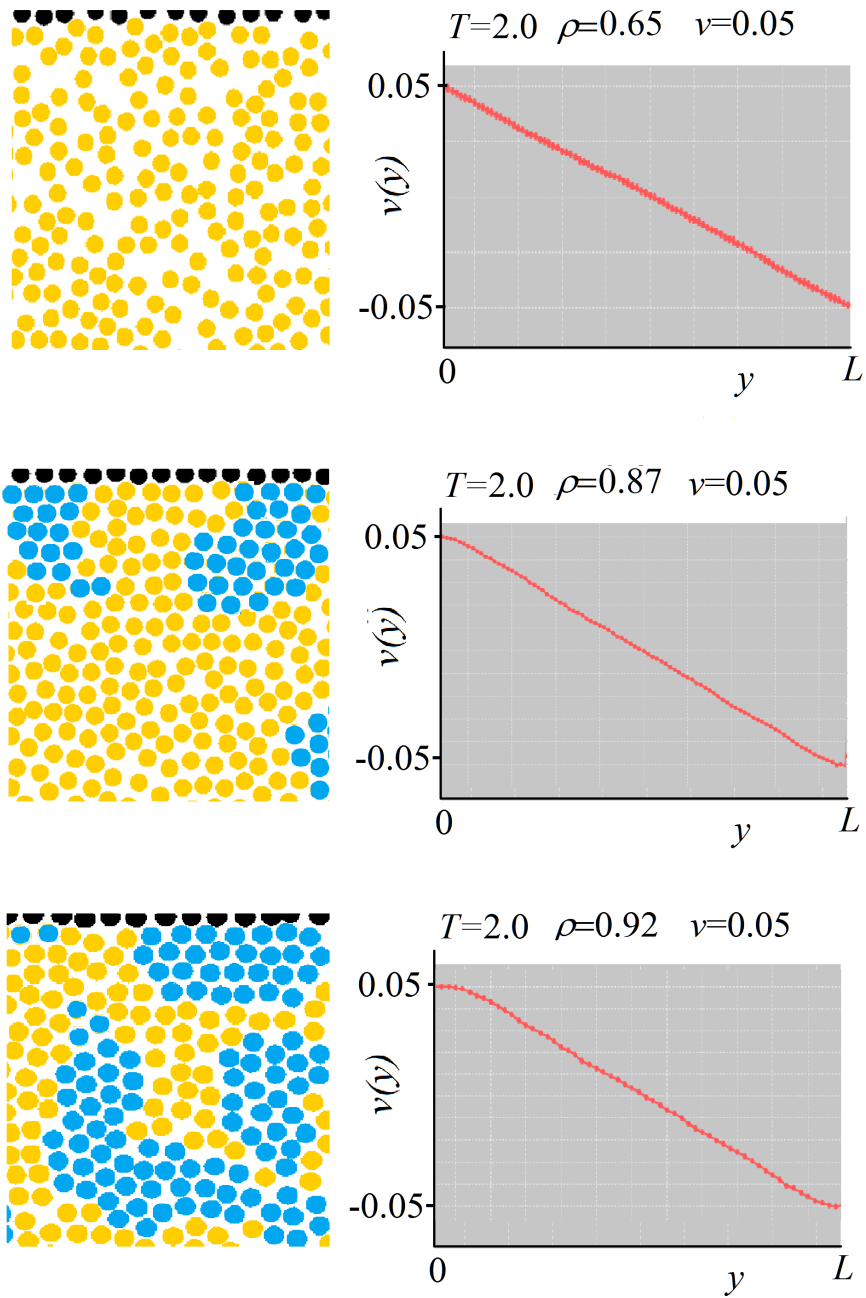

Fig. 2 (color online). The structure and the asymptotic velocity profiles in liquids at T=2 but different densities $\rho$=0.65, 0.87, and 0.92.

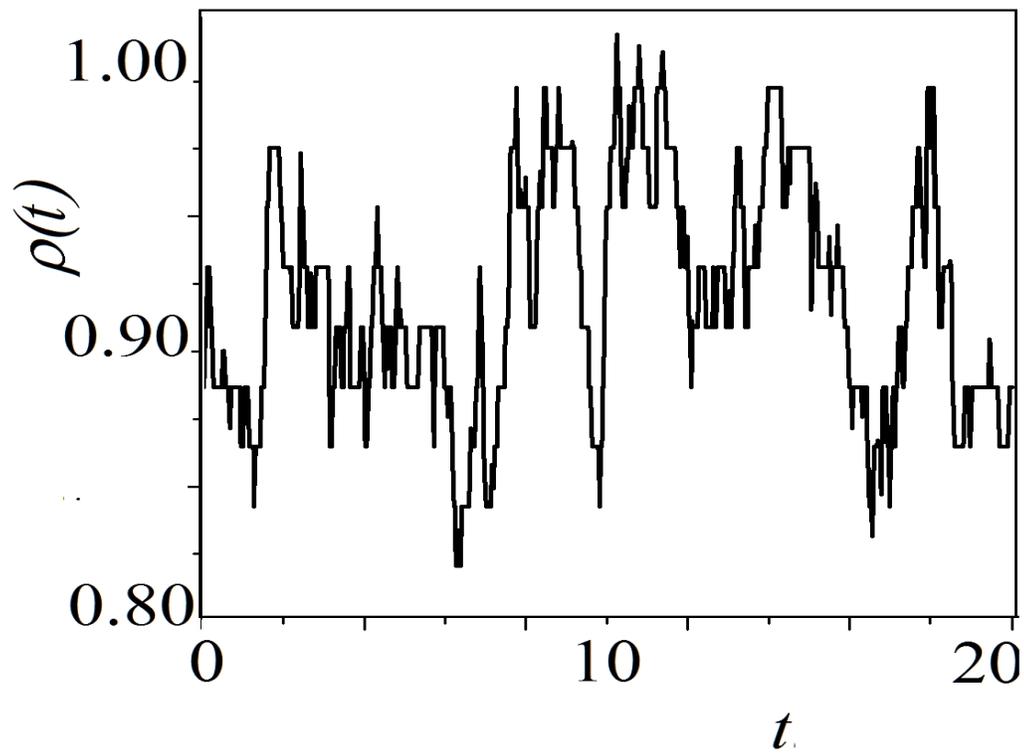

Fig. 3. Density oscillations in a 5x5 cell; sampling time $\Delta t=20$.

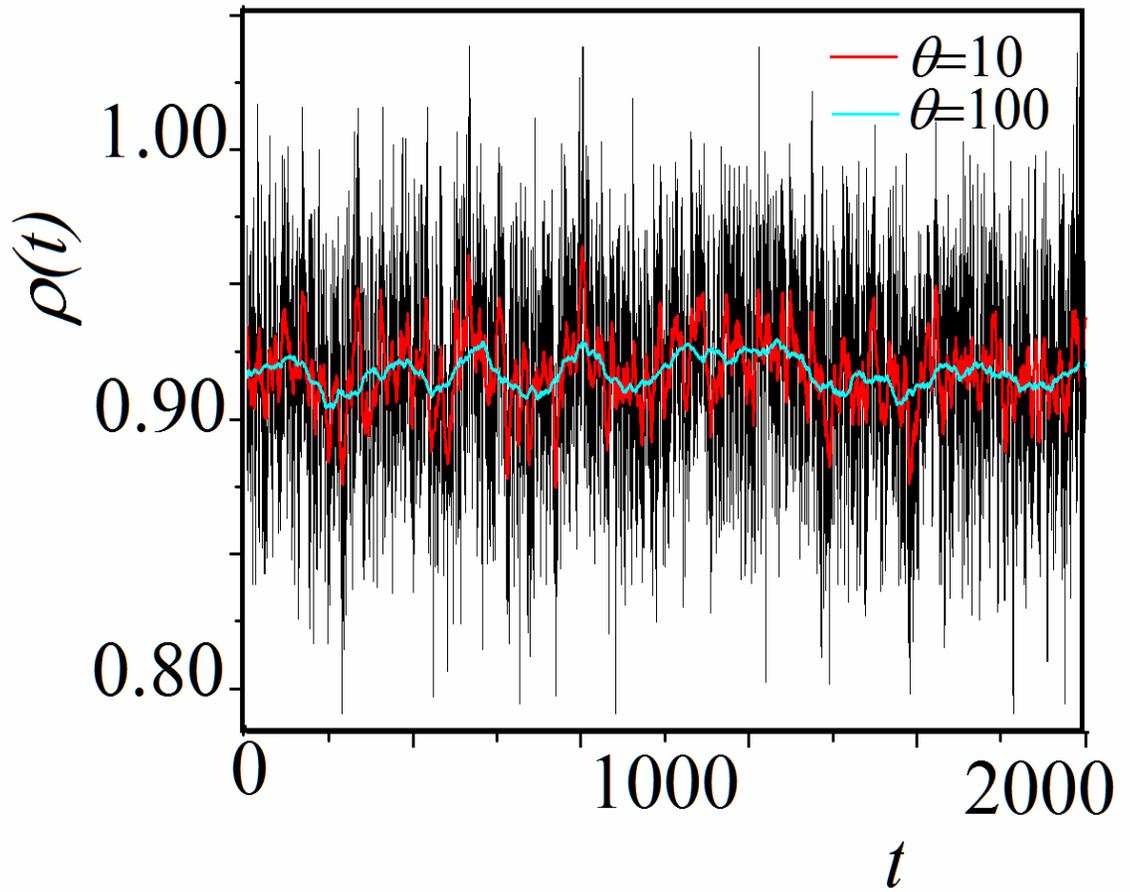

Fig. 4. $\Delta t=2000$ ($10^6 h$) sampling of the density. Smoothing times: 0.2 (100h, black line), $\theta=10$ (5000h, red line), and $\theta=100$ (50000h, cyan line).

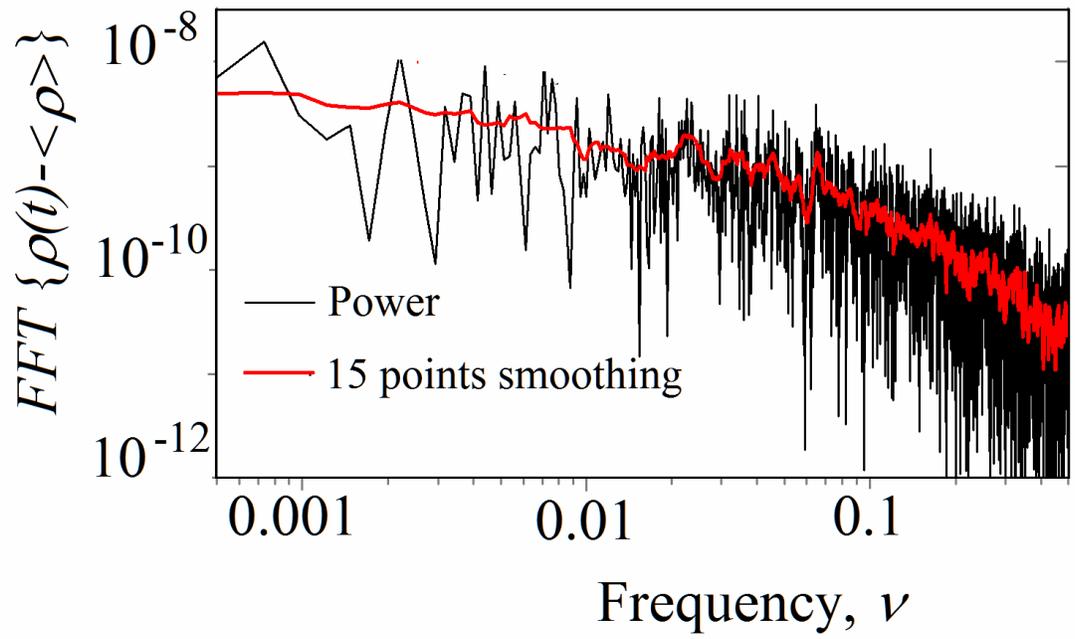

Fig. 5. *FFT* (*Fast Fourier Transform*) of density fluctuations shown in Fig. 5.

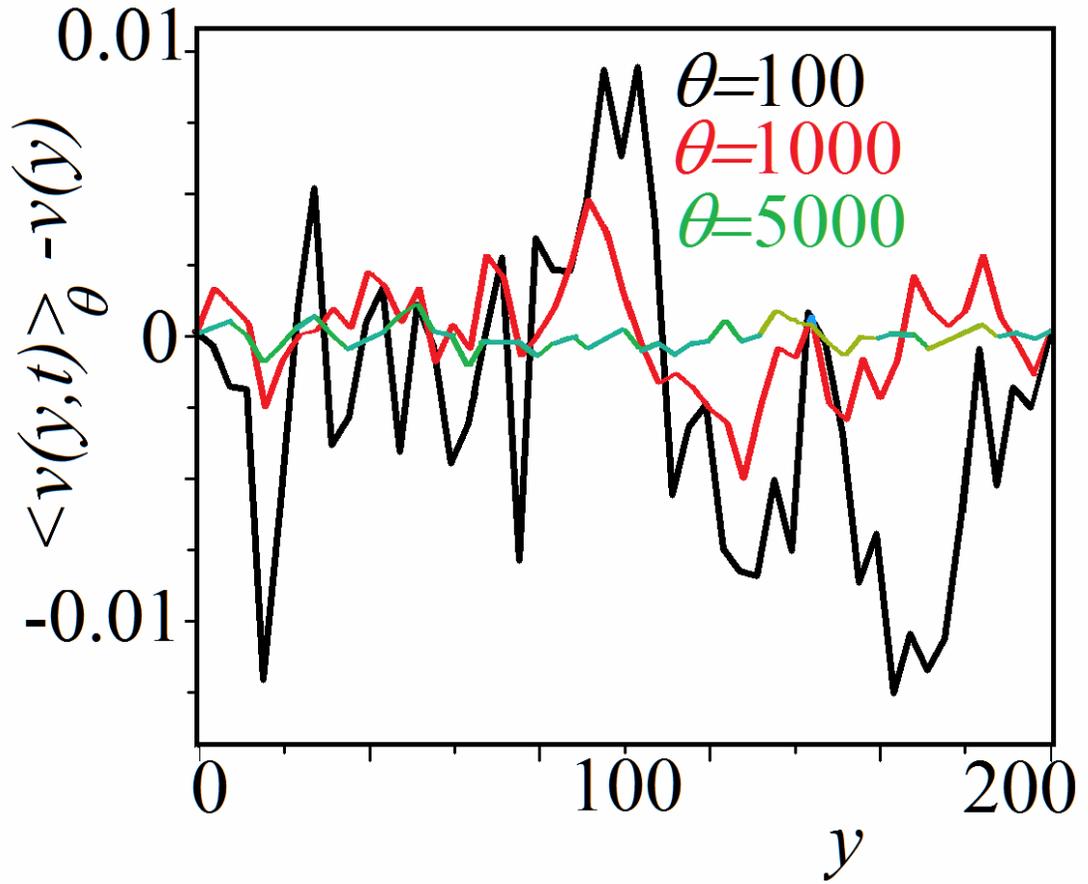

Fig. 6. Changes in $<v_x(y,t)>_\theta$ with increasing smoothing time $\theta$.

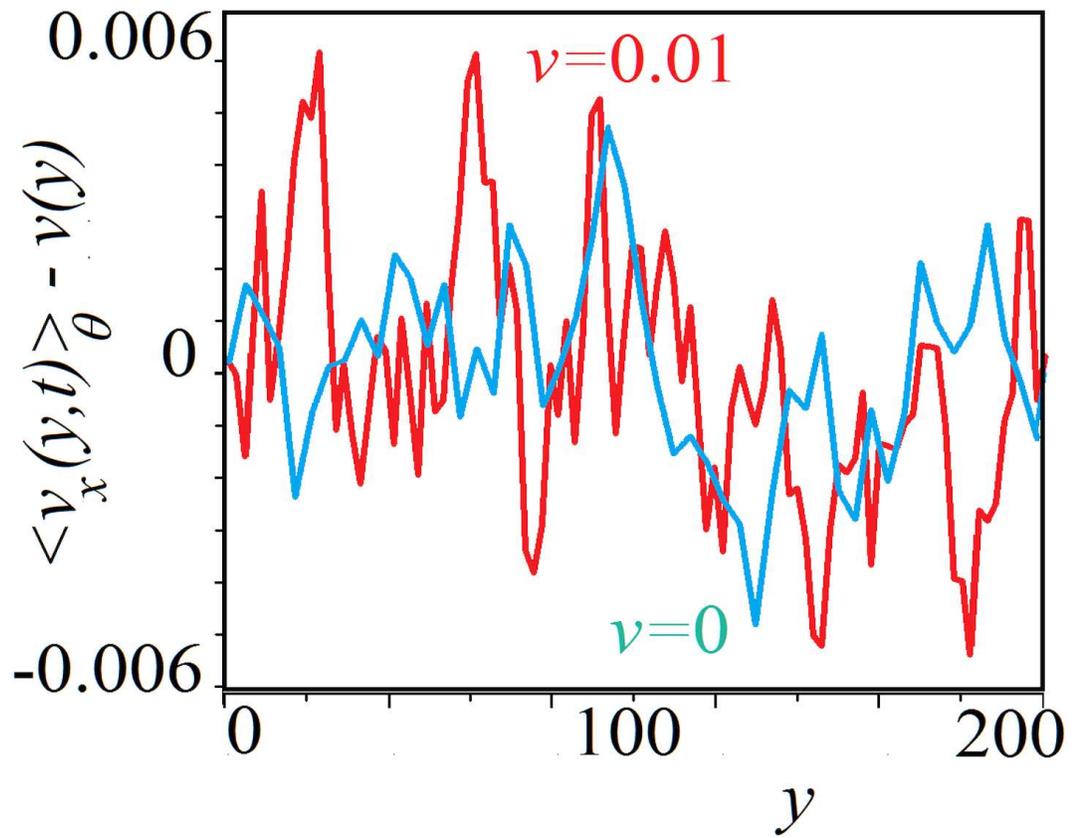

Fig. 7. Comparison of equilibrium ($v=0$, blue line) and non-equilibrium ($v=0.01$, red line) velocity $\delta\langle v_x(y,t)\rangle_\theta$ fluctuations.

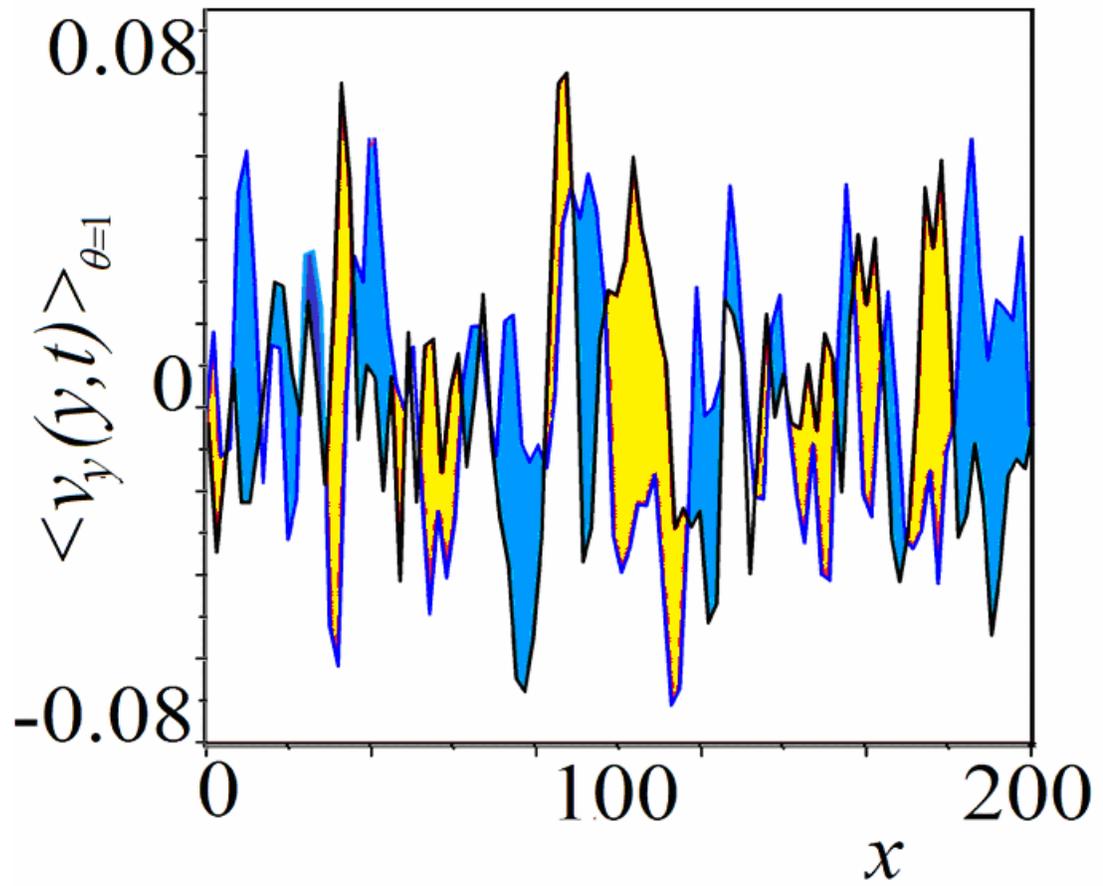

Fig. 8 (color online). Velocity profiles $<v_y(y,t)>_{\theta=1}$ at times $t$ and $t+3$. The blue (yellow) regions mark velocity increase (decrease).

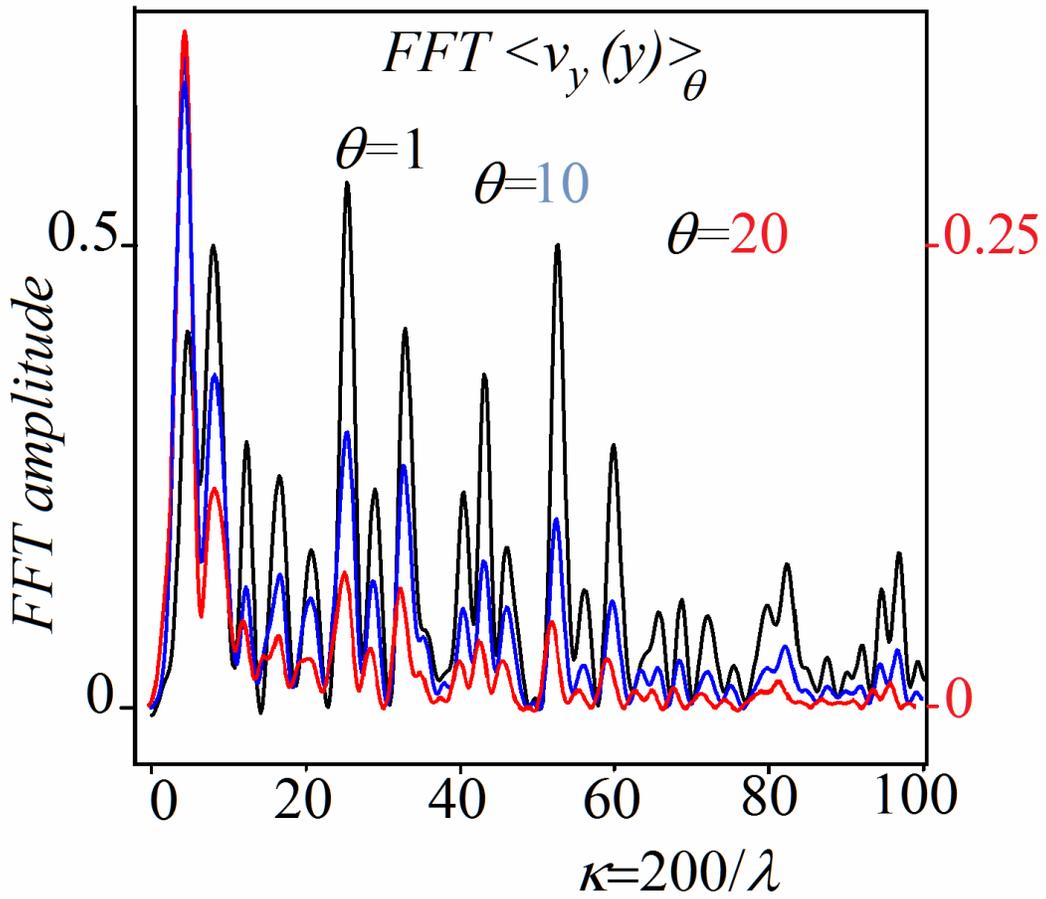

Fig. 9. *FFT*-amplitudes of smoothed velocity profiles for different smoothing times, $\theta$. Note the change of units for the red line.